\documentclass[10pt]{iopart}

\usepackage{iopams}
\usepackage{graphicx}
\usepackage{bm}
\usepackage{natbib}
\usepackage{mathrsfs}
\usepackage{dsfont}

\newcommand{\text}[1]{\mathrm{#1}}
\newcommand{\eqref}[1]{(\ref{#1})}
\renewenvironment{thebibliography}[1]{\References}{\endrefs}

\newcommand{\abs}[1]{\ensuremath{\left| #1\right|}}

\newcommand{\be}{\begin{equation}}
\newcommand{\ee}{\end{equation}}
\renewcommand{\bs}{\numparts}
\renewcommand{\es}{\endnumparts}

\newcommand{\R}{\ensuremath{\mathds{R}}}
\newcommand{\C}{\ensuremath{\mathds{C}}}
\newcommand{\N}{\ensuremath{\mathds{N}}}

\newcommand{\Th}{\ensuremath{\Theta}}
\newcommand{\Ga}{\ensuremath{\Gamma}}
\newcommand{\La}{\ensuremath{\Lambda}}

\newcommand{\usum}{\ensuremath{\sum_{n=-\infty}^\infty}}
\newcommand{\esum}{\ensuremath{\sum_{n=1}^\infty}}

\newcommand{\pd}[2][]{\ensuremath{\frac{\partial #1}{\partial #2}}}
\newcommand{\dd}[2][]{\ensuremath{\frac{\mathrm{d} #1}{\mathrm{d} #2}}}
\newcommand{\df}{\ensuremath{\mathrm{d}}}

\begin{document}

\title[Summing General Kapteyn Series]
{Methods for Summing General Kapteyn Series}

\author{R.\,C.~Tautz$^1$, I.~Lerche$^2$, and D. Dominici$^3$}

\eads{\mailto{rct@gmx.eu}, \mailto{lercheian@yahoo.com}, \mailto{dominicd@newpaltz.edu}}

\vspace{5pt}
\address{$^1$Zentrum f\"ur Astronomie und Astrophysik, Technische Universit\"at Berlin,\\
Hardenbergstra\ss{}e 36, D-10623 Berlin, Germany\\[2pt]
$^2$Institut f\"ur Geowissenschaften, Naturwissenschaftliche Fakult\"at III,\\
Martin-Luther-Universit\"at Halle, D-06099 Halle, Germany\\[2pt]
$^3$Department of Mathematics, State University of New York at New Paltz,\\
1~Hawk Dr., New Paltz, NY 12561-2443, USA}

\date{\today}

\begin{abstract}
The general features and characteristics of Kapteyn series, which are a special type of series involving Bessel function, are investigated. For many applications to physics, astrophysics, and mathematics, it is crucial to have closed-form expressions in order to determine their functional structure and parametric behavior. Closed-form expressions of Kapteyn series have mostly been limited to special cases, even though there are often similarities in the approaches used to reduce the series to analytically tractable forms. The goal of this paper is to review the previous work in the area and to show that Kapteyn series can be expressed as trigonometric or gamma function series, which can be evaluated in closed form for specific parameters. Two examples with a similar structure are given, showing the complexity of Kapteyn series.
\end{abstract}


\section{Introduction}\label{intro}

Kapteyn series of the first and second kind have arisen in a large variety of physics problems since their discovery by \citet{kap93:bes}. Series of the first kind are of the form
\be\label{eq:K1}
K_1\equiv K_1\!\left(\{a_n\},\alpha,\beta,c,b\right)=\usum a_nJ_{\alpha n+\beta}(cn+b)
\ee
with $\alpha,\beta,c$, and $b$ fixed but possibly complex, and $\{a_n\}$ is a sequence of complex coefficients. Series of the second kind are of the form
\begin{eqnarray}
K_2&\equiv K_2\!\left(\{a_n\},\alpha,\beta,\gamma,\epsilon,c,b,f,g\right)\nonumber\\
&=\usum a_nJ_{\alpha n+\beta}(cn+b)J_{\gamma n+\epsilon}(fn+g) \label{eq:K2}
\end{eqnarray}
with $\alpha,\beta,\gamma,\epsilon,c,b,f$, and $g$ fixed but possibly complex and $\{a_n\}$ is a sequence of complex coefficients. Over the years, the subject of summing Kapteyn series has attracted both physics (see below) and mathematics researchers \citep[e.\,g.,][]{erd81:trn,wat66:bes}.

One of the more important aspects of such Kapteyn series is to provide closed form expressions for the series in particular problems of physical interest. This aspect is often crucial, as one attempts to determine the functional structure and parametric behavior for the problem at hand. For example, for a system of $N$ equally charged particles uniformly spaced on a rotating ring, \citet{bud24:kap} used Kapteyn series to show how the far-field radiation distribution varies as $N\to\infty$ \citep[see also][]{ler08:kp2}. Earlier, \citet{sch12:rad} had discussed radiation from a single relativistic particle moving in a circle. Both radiation problems involved Kapteyn series of the second kind with $\alpha=\gamma=1$; $\beta=\epsilon=0$; $c=f$ and $b=g=0$ but with different coefficients $a_n$ for the two problems.

To name but a few illustrations from a broad range of physical applications \citep[see][for a review]{tau09:rev}, the large variety of physics problems involving various Kapteyn series includes: (i) Kepler's problem \citep{dat98:bes} (ii) pulsars \citep{har75:pul,ler07:kap}; (iii) side-band spectra of tera-hertz electromagnetic waves \citep{cit99:qua,ler09:thz}; (iv) high-intensity Compton scattering \citep{har09:com,ler10:com}; (v) queueing theory \citep{dom07:que}; and (vi) cosmic ray transport theories \citep{tau10:die,tau10:soq}.

In \citet{dom07:new}, Kapteyn series of the form
\be
K_1(z,t)=%
{\displaystyle\sum\limits_{n=1}^{\infty}}
t^{n}J_{n}\left(  nz\right)
\ee
were studied and a series representation was derived in powers of $z$. Furthermore, the radius of convergence was analyzed. Also, general Kapteyn series of the first kind were considered. In \citet{dom10:her}, an asymptotic approximation in terms of a Kapteyn series was obtained for the zeros of the Hermite polynomials.

As far as can be determined, physical applications of Kapteyn series to date seem to involve only structural behaviors of the form
\be
\sum a_nJ_{n+\nu}(cn)
\ee
for Kapteyn series of the first kind and
\be
\sum a_nJ_{n+\beta}(cn)J_{-(n+\epsilon)}(cn)
\ee
for Kapteyn series of the second kind. What would be of value is to determine broad ranges of the parameters and coefficients $a_n$ so that rather general closed-form representations of the Kapteyn series $K_1$ and $K_2$ are available. Such knowledge would then obviate having to evaluate each application of Kapteyn series \emph{de novo}.

Efforts in this general direction have been provided by \citet{nie01:kap,nie04:cyl} who summed particular Kapteyn series of the second type. Curiously, in respect of Nielsen's work, \citet{wat66:bes} remarks ``series of the type
\begin{equation*}
\sum\beta_nJ_{\nu+n}\left[\left(\frac{\mu+\nu}{2}+n\right)z\right]J_{\mu+n}\left[\left(\frac{\mu+\nu}{2}+n\right)z\right]
\end{equation*}
have been studied in some detail by \citet{nie01:kap}. But the only series of this type which have, as yet, proved to be of practical importance, are some special series with $\mu=\nu$, and with simple coefficients.'' However, Watson also goes on to say that ``\citet{sch12:rad} has shown that''
\begin{equation*}
\esum J_n(nz)^2=\frac{1}{2}\left[\left(1-z^2\right)^{-1/2}-1\right];
\end{equation*}
but direct inspection of equation~(31) from \citet{nie01:kap} shows that the formula ascribed to Schott was already available. Indeed many further direct summations of Kapteyn series of the second kind are to be found in Nielsen's \citeyearpar{nie01:kap} work such as [his equation~(31a)]
\be
2\esum J_n[(2n+1)x]J'_n[(2n+1)x]=\frac{1}{2x}\left[\left(1-4x^2\right)^{-1/2}-1\right].
\ee

The difference in philosophy between Nielsen and Schott is that Nielsen treated the summation as a pure mathematics' problem requiring summation, while Schott worked out the physics problem of synchrotron radiation involving the series. Thus, physics applications of the Kapteyn series arose a decade (or so) after the series was originally summed in closed form.

It would seem that to prejudge the ability to provide summations of as many as possible Kapteyn series of the second kind as not of practical use is basically not appropriate, for applications often follow much later than the basic mathematical results.

For these reasons, it seems relevant to consider \emph{de novo} the series $K_2$ and to attempt to determine procedures for evaluating such series for as broad a range of parameters and coefficients as possible. In Secs.~\ref{k2series} and \ref{exem}, general methods as well as two specific examples for the summation of Kapteyn series of the second kind will be presented, respectively. Sec.~\ref{summ} provides a short summary and a discussion of the results.

\section{Methods for Summing $K_2$ Series}\label{k2series}

In this section, different approaches are investigated that have proven useful for summing various Kapteyn series of the second kind.

\subsection{An Integral Representation Procedure}

One of the main techniques for summing Kapteyn series of the second kind is \citep{erd81:trn,gr:int}
\be\label{eq:wh}
J_\mu(z)J_\nu(z)=\frac{2}{\pi}\int_0^{\pi/2}\df\theta\;J_{\mu+\nu}(2z\cos\theta)\cos(\mu-\nu)\theta,
\ee
which is valid for $\mu,\nu$ any integer values and is otherwise valid for $\Re(\mu+\nu)>-1$. Note here the requirement that $J_\mu$ and $J_\nu$ have the same argument.

Thus, when considering
\be
K_2\equiv\usum a_n J_{\alpha n+\beta}(cn+b)J_{\gamma n+\epsilon}(fn+g)
\ee
one restricts the evaluation using equation~\eqref{eq:wh} to $c=f$ and $b=g$. Then, for $\Re(\mu+\nu)>-1$ one requires $\Re[(\alpha+\gamma)n+\beta+\epsilon]>-1$ for all integers $n$, with $a_n\neq0$. If $a_n\neq0$ for some integer $n$, then one requires $\Re(\alpha+\gamma)=0$ and $\Re(\beta+\epsilon)>-1$.

Hence, one has to evaluate
\be
K_2=\usum a_nJ_{\alpha n+\beta}(cn+b)J_{-\alpha_{\text R}n+\rmi\gamma_{\text I}n+\epsilon}(cn+b),
\ee
where $\gamma=\gamma_{\text R}+\rmi\gamma_{\text I}$ with $\gamma_{\text R}=-\alpha_{\text R}$ and $\alpha=\alpha_{\text R}+\rmi\alpha_{\text I}$, and $\Re(\beta+\epsilon)>-1$.

Starting with the case $\mu=-\nu$ in equation~\eqref{eq:wh}, one has for the Kapteyn series of the second kind, $K_2(\{a_n\},\alpha,\beta,-\alpha,-\beta,c,b,c,b)$, i.\,e.,
\bs
\begin{eqnarray}
K_2&=&\usum a_nJ_{\alpha n+\beta}(cn+b)J_{-(\alpha n+\beta)}(cn+b)\\
&=&\frac{2}{\pi}\int_0^{\pi/2}\usum a_nJ_0\!\left[2\cos\theta(cn+b)\right]\cos\!\left[2\theta(\alpha n+\beta)\right].
\end{eqnarray}
\es
Using the fact that
\be
J_0(z)=\frac{1}{\pi}\int_0^\pi\df\psi\;\cos(z\sin\psi)
\ee
one gets
\bs
\begin{eqnarray}
\fl K_2&=&\frac{2}{\pi^2}\int_0^{\pi/2}\!\df\theta\!\int_0^\pi\!\df\psi\!\usum a_n\cos\!\left[2\theta(\alpha n+\beta)\right]\cos\!\left[2\cos\theta\sin\psi(cn+b)\right]\\
\fl &=&\frac{1}{\pi^2}\int_0^{\pi/2}\df\theta\int_0^\pi\df\psi\usum a_n\nonumber\\
\fl &\times&\sum_{j=\pm1}j\left\{\cos\!\left[2\theta(\alpha n+\beta)-2j\cos\theta\sin\psi(cn+b)\right]\right\}.
\end{eqnarray}
\es
To the extent that one can sum expressions such as
\be
S=\usum a_n\cos\!\left(nA+B\right)
\ee
in closed form, $K_2$ can be, at worst, reduced to a double integral and, at best, can be evaluated in closed form. The reduction depends precisely on the functional forms chosen for $a_n$ and on the convergence of the terms in the series $S$ (normally, but not necessarily, by taking $A$ to be real).

For example, if $a_n=(n+p)^{-1}$ where $p$ is not an integer, then one can write
\be
S=-pS'\cos B+\pd[S']A\,\sin B,
\ee
where
\be
S'=\usum\frac{\cos nA}{n^2-p^2}.
\ee
Because $S'$ can be given in closed form, it is possible to sum a variety of Kapteyn series of the second kind with this procedure.

Thus, for sideband spectra in the tera-hertz regime one can show \citep{ler09:thz} that
\be
\sum_{\nu=1}^\infty\frac{J_{\nu+n}(a\nu)J_{\nu-n}(a\nu)}{\nu^2-b^2}=(-1)^{n-1}\frac{\pi}{2b}\,\csc(\pi b)J_{n+b}(ab)J_{n-b}(ab)
\ee
for $n$ integer and $n\geqslant1$, with $0<a<1$ and $0<b<1$ together with
\begin{eqnarray}
&&\sum_{\nu=0}^\infty\frac{(-1)^{n-\nu}}{\left(\nu+\frac{1}{2}\right)^2-b^2}\,J_{n+\nu+1}\!\left(a(\nu+{\textstyle\frac{1}{2}})\right)J_{n-\nu}\!\left(a(\nu+{\textstyle\frac{1}{2}})\right)\nonumber\\
=&&\;(-1)^n\frac{\pi}{4b}\,\sec(\pi b)J_{n+\frac{1}{2}+b}(ab)J_{n+\frac{1}{2}-b}(ab)
\end{eqnarray}
for $n$ integer and $n\geqslant1$, with $0<a<1$ and $0<b<\frac{1}{2}$.

Similarly, for high-intensity Compton scattering, series arise of the form
\be
\esum\frac{n^p}{(a+n)^q}\,J_n^2(na)
\ee
with $p$ and $q$ positive integers, which can be reduced to analytic closed form apart from a single elliptic integral that has to be added to the rest of the closed form expressions \citep{ler10:com}.

Besides such evaluations where $\mu=-\nu$, the more general case with $\Re(\mu+\nu)>-1$ needs to be considered. Writing equation~\eqref{eq:wh} with
\bs
\begin{eqnarray}
\mu&=&\alpha n+\beta\\
\nu&=&-\alpha_{\text R}n+\rmi\gamma_{\text I}n+\epsilon,
\end{eqnarray}
\es
one obtains
\bs
\begin{eqnarray}
\mu+\nu&=&\rmi(\gamma_{\text I}+\alpha_{\text I})n+\beta+\epsilon\\
\mu-\nu&=&2\alpha_{\text R}n+\rmi(\alpha_{\text I}-\gamma_{\text I})n+\beta-\epsilon.
\end{eqnarray}
\es
Thus, one can write
\begin{eqnarray}
K_2&=&\frac{2}{\pi}\int_0^{\pi/2}\df\theta\usum a_nJ_{in(\gamma_{\text I}+\alpha_{\text I})+\beta+\epsilon}\bigl(2\cos\theta(cn+b)\bigr)\nonumber\\
&\times&\cos\!\left[\theta\bigl(2\alpha_{\text R}n+\rmi(\alpha_{\text I}-\gamma_{\text I})n+\beta-\epsilon\bigr)\right] \label{eq:K2a}
\end{eqnarray}
with $\Re(\beta+\epsilon)>-1$. The cosine factors in equation~\eqref{eq:K2a} converge if and only if $\alpha_{\text I}=\gamma_{\text I}$, so that
\be\label{eq:K2b}
\fl K_2=\frac{2}{\pi}\int_0^{\pi/2}\df\theta\usum a_nJ_{2\rmi n\alpha_{\text I}+\beta+\epsilon}\bigl(2\cos\theta(cn+b)\bigr)\cos\left[\theta(2\alpha_{\text R}n+\beta-\epsilon\right].
\ee
Therefore,
\be
J_\nu(x)=\frac{2}{\pi}\int_0^x\df t\;\sin\left(x\cosh t-\frac{\pi\nu}{2}\right)\cosh\nu t,
\ee
for $x\in\R$, so that $c$ and $b$ are real. But if $\nu$ contains an imaginary part proportional to $n$, as in equation~\eqref{eq:K2b}, then the series in equation~\eqref{eq:K2b} diverges exponentially unless $a_n$ converges fast enough (e.\,g., $a_n$ proportional to $\exp[-bn^2]$). Thus, one requires $\alpha_{\text I}\equiv0$.

Under such conditions one can write
\begin{eqnarray}
K_2&=&\left(\frac{2}{\pi}\right)^2\int_0^\infty\df t\int_0^{\pi/2}\df\theta\usum a_n\cos\!\left[\nu(2\alpha_{\text R}n+\beta-\epsilon)\right]\nonumber\\
&\times&\cosh\bigl[(\beta+\epsilon)t\bigr]\sin\left[2\cos\theta(cn+b)\cosh t-\frac{\pi}{2}\left(\beta+\epsilon\right)\right]. \label{eq:K2c}
\end{eqnarray}
Again, one sees that for choices of $a_n$ such that the series in equation~\eqref{eq:K2c} can be summed in closed form then $K_2$ is reduced, at worst, to a double integral and, at best, can be evaluated explicitly.

All of these procedures for summing the general second-order Kapteyn series represented by $K_2$ are dependent on the integral representation from equation~\eqref{eq:wh} for $J_\nu(z)J_\mu(z)$, valid for $\Re(\mu+\nu)>-1$ when $\mu,\nu$ are not integer and otherwise generally valid.

But just because there are values of $\Re(\mu+\nu)\leqslant-1$, for which equation~\eqref{eq:wh} cannot be used, does not mean that other Kapteyn series of the $K_2$ form cannot be summed. One needs other procedures to effect the summations when $\Re(\mu+\nu)\leqslant-1$.

\subsection{Series Representation Procedures}

\begin{figure}[tb]
\centering
\includegraphics[width=90mm]{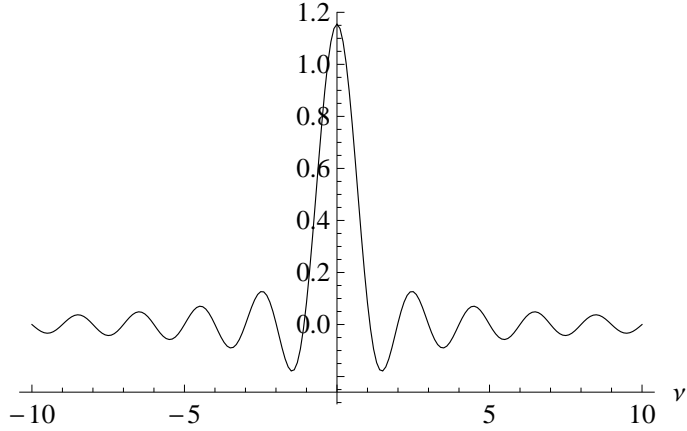}
\caption{The Kapteyn series from equation~\eqref{eq:nie2a} for varying real $\nu\in[-10,10]$ and for $x=1/2$.}
\label{ab:Nielsen1}
\end{figure}

\begin{figure}[tb]
\centering
\includegraphics[width=90mm]{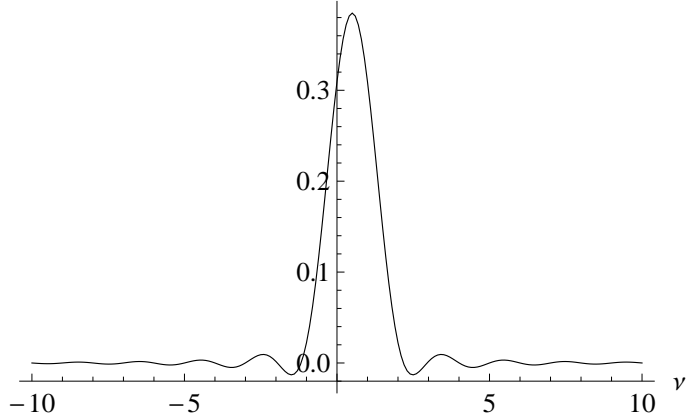}
\caption{The Kapteyn series from equation~\eqref{eq:nie2b} for varying real $\nu\in[-10,10]$ and for $x=1/2$.}
\label{ab:Nielsen2}
\end{figure}

The use of series representations to sum $K_2$ types of Kapteyn series was already known to \citet{nie01:kap} and later the same procedure was given by \citet{wat66:bes}. Following \citet{nie01:kap}, the sense of the argument is as follows: One considers first integrals of the form
\be
\mathcal I_\nu(a)=\int_0^{\pi/2}\df x\;\cos^{\nu-1}x\cos ax
\ee
and, for $\Re(\nu)>0$, expresses the result as
\be
\mathcal I_\nu(a)=\frac{\pi\cos a\pi/2}{2^\nu\nu}\,\frac{\Ga(\nu+1)}{\Ga\bigl((\nu+1+a)/2)\bigr)\Ga\bigl((\nu+1-a)/2\bigr)}.
\ee
Then one considers equation~\eqref{eq:wh} with $\mu=n-a$ and $\nu=n+a$. According to Nielsen, one then uses the series expansion definition of the Bessel function under the integral sign as
\be\label{eq:ser}
J_\alpha(z)=\left(\frac{z}{2}\right)^\alpha\sum_{k=0}^\infty\frac{(-1)^k}{2^{2k}k!}\,\frac{z^{2k}}{\Ga(\alpha+k+1)}
\ee
and so one integrates equation~\eqref{eq:wh} term by term with $\mu$ and $\nu$ as defined above by using equation~\eqref{eq:ser}. Effectively, one trades a sum over Bessel functions of the Kapteyn kind for a power series. The resulting power series can often, but not universally, be either summed in closed form or can be evaluated for specific parameter values. In this way, Nielsen argued that
\be\label{eq:nie2a}
\fl \frac{\sin\nu\pi}{\nu\pi}+2\esum J_{n+\nu}(2nx)J_{n-\nu}(2nx)=\frac{1}{\sqrt\pi}\sum_{n=0}^\infty\frac{n!\,\Ga(n+\frac{1}{2})(2x)^{2n}}{\Ga(n+1+\nu)\Ga(n+1-\nu)},
\ee
and
\be\label{eq:nie2b}
\fl 2\sum_{n=0}^\infty J_{n+\nu}[(2n+1)x]J_{n+1-\nu}[(2n+1)x]=\frac{1}{\sqrt\pi}\sum_{n=0}^\infty\frac{n!\,\Ga(n+\frac{3}{2})(2x)^{2n+1}}{\Ga(n+1+\nu)\Ga(n+2-\nu)},
\ee
which are illustrated in figures~\ref{ab:Nielsen1} and \ref{ab:Nielsen2}, respectively, for varying $\nu$ and for $x=1/2$. Numerically, the evaluation of infinite series is carried out as follows: First, a number of terms (usually $100$) is summed directly; to accelerate the convergence of the sum, then for example Wynn's epsilon method \citep[e.\,g.,][]{bre00:acc,ham86:num} can be used, which samples a number of additional terms (usually $100$) in the sum and tries to fit a polynomial multiplied by a decaying exponential. Thus, the series are well approximated and the required computer time is kept moderate.

Note that both equations~\eqref{eq:nie2a} and \eqref{eq:nie2b} are valid for arbitrary $\nu\in\C$ and for complex $x\in\C$ with $\abs x<1/2$. However, for $\Im(\nu)>1$, both series attain extremely large values, depending on $\Re(\nu)$ and $x$.

Nielsen then commented that for $\nu=0$ one has the particular cases
\be\label{eq:nie_res1}
1+2\esum J_n(2nx)^2=\left(1-4x^2\right)^{-1/2}
\ee
and
\be\label{eq:nie_res2}
2\sum_{n=0}^\infty J_n[(2n+1)x]J_{n+1}[(2n+1)x]=\frac{1}{2x}\left[\left(1-4x^2\right)^{-1/2}-1\right].
\ee
Note that the results shown here correct two misprints in Nielsen's results [his equations~(30a) and (31)], which are: (i) Nielsen wrote $J_{n-\nu}[(2n+1)x]$ on the left-hand side of equation~\eqref{eq:nie2b} instead of $J_{n+1-\nu}[(2n+1)x]$ and (ii) Nielsen included the term $n=0$ when summing the left-hand side of equation~\eqref{eq:nie_res1}.

In fact, however, a more general representation is possible if, instead of using Nielsen's series expansion of the Bessel function under the integral sign of equation~\eqref{eq:wh}, one were to write
\be
J_{2n}(2nx\cos\theta)=\frac{1}{\pi}\int_0^\pi\df\psi\;\cos2n\psi\cos(2nx\cos\psi)
\ee
then one sees immediately that the ability to evaluate a Kapteyn series is again reduced to the question of whether one can sum
\begin{equation*}
\esum\alpha_n\cos2n\psi\cos(2nx\cos\psi)
\end{equation*}
for particular choices of $\alpha_n$. Nielsen's choice of $\alpha_n=1$ is just one example where the summation can be achieved.

Perhaps of more general interest is to ask how reducible equation~\eqref{eq:K2} is for arbitrary parameter values. Then, using equation~\eqref{eq:ser} for \emph{each} of the Bessel functions in $K_2$ one has
\begin{eqnarray}
K_2&=&\usum a_n\sum_{k=0}^\infty\sum_{r=0}^\infty\left(\frac{cn+b}{2}\right)^{\alpha n+\beta}\frac{(-1)^k(cn+b)^{2k}}{2^{2k}k!\,\Ga(\alpha n+\beta+k+1)}\nonumber\\
&\times&\left(\frac{fn+g)}{2}\right)^{\gamma n+\epsilon}\frac{(-1)^r(fn+g)^{2r}}{2^{2r}r!\,\Ga(\gamma n+\epsilon+r+1)}. \label{eq:K2gen}
\end{eqnarray}
Unless $fn+g=\La(cn+b)$, where \La\ is a constant, it is difficult to make further headway with equation~\eqref{eq:K2gen}. But when such is the case then one has
\begin{eqnarray}
K_2&=&\usum B_n\sum_{k=0}^\infty\sum_{r=0}^\infty\left(\frac{cn+b}{2}\right)^{(\alpha+\gamma)n+\epsilon+\beta}\nonumber\\
&\times&\frac{(-1)^k(-1)^r}{2^{2k}2^{2r}k!r!}\,\frac{(cn+b)^{2(k+r)}\La^{2r}}{\Ga(\alpha n+\beta+k+1)\Ga(\gamma n+\epsilon+r+1)},
\end{eqnarray}
where $B_n=a_n\La^{\gamma n+\epsilon}$. Now, setting $k+r=m$, one gets
\begin{eqnarray}
K_2&=&\usum B_n\left(\frac{cn+b}{2}\right)^{(\alpha+\gamma)n+\epsilon+\beta}\sum_{m=0}^\infty\sum_{k=0}^m\left(\frac{cn+b}{2}\right)^{2m}\nonumber\\
&\times&\frac{(-1)^m\La^{2m}\La^{-2k}}{k!(m-k)!\,\Ga(\alpha n+\beta+k+1)\Ga(\gamma n+\epsilon+1+m-k)}. \label{eq:K2ser}
\end{eqnarray}
The representation of $K_2$ in closed form (or at worst as an integral) then rests on the extent to which one can sum the various component sums occurring in equation~\eqref{eq:K2ser}.

One can write
\be
K_2=\usum B_n\left(\frac{cn+b}{2}\right)^{(\gamma+\alpha)n+\epsilon+\beta}Q_n,
\ee
where
\begin{eqnarray}
Q_n&=&\sum_{k=0}^m\sum_{m=0}^\infty\left(\frac{cn+b}{2}\right)^{2m}\nonumber\\
&\times&\frac{(-1)^m\La^{2m}\La^{-2k}}{k!(m-k)!\,\Ga(\alpha n+\beta+k+1)\Ga(\gamma n+\epsilon+1+m-k)}.
\end{eqnarray}
Note that the replacement of $r$ by $(m-k)$ causes $Q_n$ to be a single power series in $(cn+b)/2$ which can be written as
\be
Q_n=\sum_{m=0}^\infty\left(\frac{cn+b}{2}\right)^{2m}(-1)^m\La^{2m}R_m
\ee
with
\be
\fl R_m=\sum_{k=0}^\infty\frac{\La^{2k}}{k!(m-k)!}\bigl[\Ga\!\left(\alpha n+\beta+k+1\right)\Ga\!\left(\gamma n+\epsilon+(m-k)+1\right)\bigr]^{-1}.
\ee
The basic question is: under what conditions is $R_m$ expressible in closed form? If it is, then one can then determine the conditions under which $Q_n$ is expressible in closed form and so arrange values of $B_n$ so that $K_2$ is in closed form. It would seem that only for particular values of the parameters it is possible to effect closed-form results, as those for instance given by Nielsen [his equation~(8)]
\begin{eqnarray}
\fl \left(\frac{x}{2}\right)^{\mu+\nu}&=&\left(\mu+\nu\right)\Ga(1+\mu)\Ga(1+\nu)\sum_{n=0}^\infty{{\mu+\nu+n-1}\choose n}\left(\mu+\nu+2n\right)^{-(\mu+\nu+1)}\nonumber\\
\fl &\times&J_{\mu+n}\!\left[(\mu+\nu+2n)x\right]J_{\nu+n}\left[(\mu+\nu+2n)x\right]
\end{eqnarray}

\section{Examples for $K_2$ series}\label{exem}

There are special cases of Kapteyn series of the second kind, which are often needed and which rely on other methods than those described above. One example is a Kapteyn series, which consists of $J_n^2(nz)$ for $z\in\C$ with $\abs z<1$, combined with a power of $n$ in the form
\be\label{eq:K2ex1}
K_2\equiv K_2\!\left(n^{2q},1,0,1,0,z,0,z,0\right)=\esum n^{2q}J_n^2(nz).
\ee

Two different distinctions can be made: (i) $q<0$; (ii) $q>0$; each for (a) integer $q\in\N$; (b) arbitrary $q\in\R$.

\subsection{The case $q<0$}

First, write $p=-q$ so that
\be
K_2=\esum\frac{J_n^2(nz)}{n^{2p}}.
\ee
The general procedure is the following: use Bessel's equation $J_n(z),$ to show that \citep[cf.][equation~17.33]{wat66:bes}, for consecutive indices $p$, the following two Kapteyn series of the first kind are related through the equation
\be\label{diffeq}
\left(z\,\dd z\right)^2\esum\frac{J_{2n}(2nz)}{n^{2p}}=4\left(1-z^2\right)\esum\frac{J_{2n}(2nz)}{n^{2(p-1)}}.
\ee
The second initial condition, i.\,e., the sum for $p=1$ \citep[cf.][Sec.~17.23]{wat66:bes},
\be
\esum\frac{J_{2n}(2nz)}{n^2}=\frac{z^2}{2},
\ee
together with Meissel's \citeyearpar{mei92:bes} investigation suggests one should write the Kapteyn series of the first kind as a polynomial in $z^{2k}$ \citep[see][Sec.~17.23]{wat66:bes}. By evaluating the recurrence relation, equation~\eqref{diffeq}, it has been shown \citep{tau10:die} that the Kapteyn series of the first kind can be expressed as
\be\label{eq:Fs2}
\esum\frac{J_{2n}(2nz)}{n^{2p}}=\sum_{k=1}^pz^{2k}\sum_{j=1}^k\frac{(-1)^{j+k}j^{2(k-p)}}{(k-j)!(k+j)!}.
\ee

To obtain the corresponding Kapteyn series of the second kind, equation~\eqref{eq:K2ex1}, one employs equation~\eqref{eq:wh} for $\mu=\nu=n$. Then one evaluates equation~\eqref{eq:Fs2} with the argument $2nz\cos\theta$ and integrates over $\theta$, noting that \citep[cf.][Sec.~3.621]{gr:int}
\be\label{eq:dfac}
\frac{2}{\pi}\int_0^{\pi/2}\df\theta\;\cos^{2n}\theta=\frac{(2n-1)!!}{(2n)!!}=\frac{\Ga\!\left(n+\frac{1}{2}\right)}{n!\sqrt\pi},\qquad n\in\N,
\ee
where $(\cdot)!!$ is the double factorial and where $\Ga(\cdot)$ denotes the Gamma function. Hence, the result is
\be\label{eq:resG}
K_2(z)=\sum_{k=1}^\Th z^{2k}\,\frac{\Ga\!\left(k+\frac{1}{2}\right)}{k!\sqrt\pi}\sum_{j=0}^{k-1}\frac{(-1)^j(k-j)^{2(k-p)}}{j!(2k-j)!},
\ee
with
\be
\Th=
\left\{
\begin{tabular}{ll}
\;$p$,& $p\in\N$\\
\;$\infty,$& $p\in\R\!\setminus\!\N$
\end{tabular}
\right.
\ee
and is valid for arbitrary $p\in\R<0$.

Equation~\eqref{eq:resG} has some important features. For integer $p$, a sum with $p(p+1)/2$ terms is obtained, whereas, for non-integer $p$, an infinite power series occurs. But even in that case, it is advantageous to exchange one infinite series (the Kapteyn series) by another infinite series (a power series), because the convergence behavior of a power series is better understood, thus allowing for a more reliable estimate of the number of terms needed to obtain a desired accuracy.

\begin{figure}[tb]
\centering
\includegraphics[width=100mm]{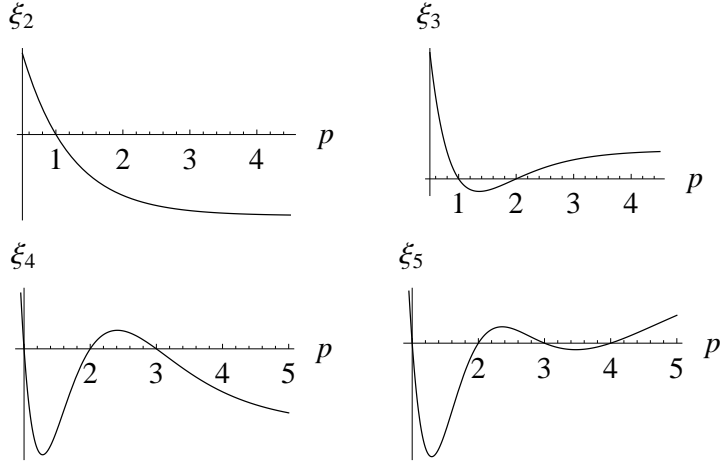}
\caption{The functions $\xi_k(p)$ for varying $p$ with $k\in\{2,3,4,5\}$ as given through equation~\eqref{eq:resG} and, in explicit form, equation~\eqref{eq:xifunctions}. The vertical axis should merely illustrate the zeros of the functions $\xi_k(p)$, thus explaining why, for $p$ integer, a \emph{finite} power series is obtained in equation~\eqref{eq:resG}.}
\label{ab:xifunctions}
\end{figure}

The reason for the distinction between a finite/infinite power series is that, when expanding the coefficients of the powers $z^{2k}$, one finds non-algebraic functions $\xi_k(p)$ of the form
\begin{eqnarray}
\xi_2(p)&=&-\frac{1}{16}+4^{-(1+p)}\nonumber\\
\xi_3(p)&=&\frac{1}{768}\left(5-2^{7-2p}+3^{5-2p}\right)\nonumber\\
\xi_4(p)&=&\frac{1}{18\,432}\left(-7+2^{13-4p}+7\cdot2^{7-2p}-9^{4-p}\right)\nonumber\\
\xi_5(p)&=&\frac{1}{2\,949\,120}\left(42-2^{21-4p}-3\cdot2^{13-2p}+5^{9-2p}+9^{6-p}\right), \label{eq:xifunctions}
\end{eqnarray}
each of which has zeros at the first integers, i.\,e., $\xi_k=0$ for $p=1,\dots,k-1$. The functions $\xi_k(p)$ are illustrated in figure~\ref{ab:xifunctions} for varying $p$ and for $k\in\{2,3,4,5\}$.

\subsection{The case $q>0$}

Even though structurally similar, the opposite case with $q>0$ results in a completely different behavior when evaluating the Kapteyn series in terms of a power series. In \citet{dom11:tay}, Kapteyn series of the form%
\be
K_{2}(z,q)\equiv K_2\!\left(n^{2q},1,0,1,0,2z,0,2z,0\right)=%
{\displaystyle\sum\limits_{n=1}^{\infty}}
n^{2q}J_{n}^{2}\left(  2nz\right)  ,
\ee
were investigated and it was found that
\begin{equation}
K_{2}(z,q)=%
{\displaystyle\sum\limits_{n=1}^{\infty}}
\left[  \frac{1}{\left(  n!\right)  ^{2}}%
{\,\displaystyle\sum\limits_{k=0}^{n}}
\left(  -1\right)  ^{k}{2n\choose k}\left(  n-k\right)  ^{2\left(  n+q\right)
}\right]  z^{2n}, \label{K2}%
\end{equation}
for $q\geqslant0$ and $\left\vert z\right\vert <1/2.$ The result in equation~(\ref{K2}) allows one to compute $K_{2}(z,q)$ numerically for any $q\geqslant0,$; however, sometimes it is difficult to use it to get closed-form expressions. Here, a different approach will be introduced using differential operators.

From \citet[equation~5.4~(4)]{wat66:bes}, it is known that the function $y_{n}(t)=J_{n}%
^{2}\left(  \rme^{t}\right)  $ satisfies the differential equation%
\begin{equation}
\frac{\df^{3}y_{n}}{\df t^{3}}+4\left(  \rme^{2t}-n^{2}\right)  \frac{\df y_{n}}%
{\df t}+4\rme^{2t}y_{n}=0. \label{eq1}%
\end{equation}
By changing variables to $2nz=\rme^{t}$ in equation~(\ref{eq1}), one obtains
\be
\fl \left[  z^{2}\frac{\df^{3}}{\df z^{3}}+3z\frac{\df^{2}}{\df z^{2}}+\left(  16n^{2}%
z^{2}-4n^{2}+1\right)  \frac{\df}{\df z}+16n^{2}z\right]  J_{n}^{2}\left(
2nz\right)  =0,
\ee
or
\be
\fl \left(  z^{2}\frac{\df^{3}}{\df z^{3}}+3z\frac{\df^{2}}{\df z^{2}}+\frac{\df}{\df z}\right)
J_{n}^{2}\left(  2nz\right)  =\left[  4\left(  1-4z^{2}\right)  \frac{\df}%
{\df z}-16\right]  n^{2}J_{n}^{2}\left(  2nz\right)  .
\ee
Introducing the function
\be
g_{q}\left(  z,n\right)  =n^{2q}J_{n}^{2}\left(  2nz\right)  ,
\ee
one has%
\be
\left(  z^{2}\frac{\df^{3}}{\df z^{3}}+3z\frac{\df^{2}}{\df z^{2}}+\frac{\df}{\df z}\right)
g_{q}=\left[  4\left(  1-4z^{2}\right)  \frac{\df}{\df z}-16\right]  g_{q+1},
\ee
for all $n\in\N^\star$, i.\,e., all positive integers.

Thus, it follows that the function $K_{2}(z,q)$ satisfies%
\begin{equation}
\fl \left(  z^{2}\frac{\df^{3}}{\df z^{3}}+3z\frac{\df^{2}}{\df z^{2}}+\frac{\df}{\df z}\right)
K_{2}(z,q)=\left[  4\left(  1-4z^{2}\right)  \frac{\df}{\df z}-16\right]
K_{2}(z,q+1), \label{EqK2}%
\end{equation}
while equation~\eqref{eq:K2} gives for the initial condition
\begin{equation}
K_{2}(z,0)=-\frac{1}{2}+\frac{1}{2\sqrt{1-4z^{2}}}. \label{K20}%
\end{equation}
Since $J_{n}\left(  0\right)  =0$ for all $n=1,2,\ldots,$ one has
$K_{2}(0,q)=0$. Solving equation~(\ref{EqK2}) for $K_{2}(z,q+1),$ one obtains%
\begin{equation}
\fl K_{2}(z,q+1)=\frac{1}{4\sqrt{1-4z^{2}}}%
{\displaystyle\int\limits_{0}^{z}}
\frac{\df w}{\sqrt{1-4w^{2}}}\left(  w^{2}\frac{\df^{3}}{\df w^{3}}+3w\frac{\df^{2}%
}{\df w^{2}}+\frac{\df}{\df w}\right)  K_{2}(w,q), \label{K2int}%
\end{equation}
where one must be careful that $\left\vert z\right\vert <1/2$ to ensure that the
integral is convergent.

\begin{figure}[tb]
\centering
\includegraphics[width=90mm]{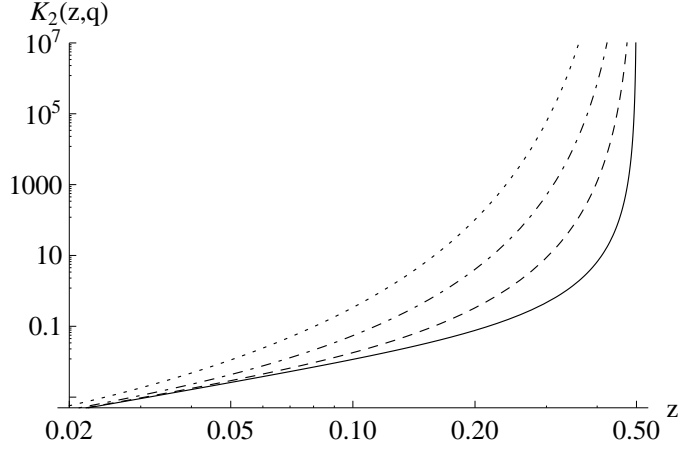}
\caption{The first functions $K_2(z,q)$ for varying $z\in[0,1/2]$ with $q\in\{1,2,3,4\}$ as given through equation~\eqref{eq:K2Dgen} and, in explicit form, equation~\eqref{eq:K2Dexpl}. The order $q$ varies from right to left, i.\,e., the solid line shows $K_2(1,z)$ while the dotted line shows $K_2(4,z)$.}
\label{ab:Diego}
\end{figure}

Integrating equation~(\ref{K2int}) by parts, one obtains the final result for the series $K_2$ in the form of a the recurrence relation, which reads
\begin{eqnarray}
\fl K_{2}(z,q+1) &  =&\left[  \frac{z^{2}}{4\left(  1-4z^{2}\right)  }\frac{\df^{2}%
}{\df z^{2}}+\frac{z\left(  1-8z^{2}\right)  }{4\left(  1-4z^{2}\right)  ^{2}%
}\frac{\df}{\df z}+\frac{2z^{2}\left(  1+2z^{2}\right)  }{\left(  1-4z^{2}\right)
^{3}}\right]  K_{2}(z,q)\label{K2int1}\nonumber\\
\fl &  -&\frac{4}{\sqrt{1-4z^{2}}}%
{\displaystyle\int\limits_{0}^{z}}
\df w\;\frac{w\left(  1+10w^{2}+4w^{4}\right)  }{\left(  1-4w^{2}\right)  ^{7/2}}\,K_{2}(w,q).
\end{eqnarray}
Using equation~\eqref{K2int1} and \eqref{K20}, it is straightforward (albeit tedious) to compute the explicit expressions for the first orders of the series $K_2$, yielding the relations%
\begin{eqnarray}
K_{2}(z,1)&=&\frac{z^{2}\left(1+z^{2}\right)}{\left(1-4z^{2}\right)^{7/2}}\nonumber\\
K_{2}(z,2)&=&\frac{z^{2}\left(1+37z^{2}+118z^{4}+27z^{6}\right)}{\left(1-4z^{2}\right)^{13/2}}\nonumber\\
K_{2}(z,3)&=&\frac{z^{2}\left(1+217z^{2}+5036z^{4}+23\,630z^{6}+22\,910z^{8}+2250z^{10}\right)}{\left(1-4z^{2}\right)^{19/2}}\nonumber
\end{eqnarray}
\begin{eqnarray}
\fl K_{2}(z,4)=z^{2}\left(385\,875z^{14}+7\,119\,756z^{12}+15\,359\,862z^{10}+8\,635\,578z^{8}\right.\nonumber\\
+\left.1\,515\,705z^{6}+80130z^{4}+973z^{2}+1\right)\left(1-4z^{2}\right)^{-25/2}, \label{eq:K2Dexpl}
\end{eqnarray}
which are illustrated in figure~\ref{ab:Diego}.

In general, one has
\be\label{eq:K2Dgen}
K_{2}(z,q)=\frac{z^{2}P_{q}\left(z^{2}\right)}{\left(1-4z^{2}\right)^{3q+1/2}},\qquad\left\vert z\right\vert <\frac{1}{2},
\ee
where $P_{q}\left(  z\right)  $ is a polynomial of degree $2q-1$. The structure of the polynomials $P_{q}(z)$ is quite complicated and will be analyzed in a forthcoming paper.

\section{Discussion and Conclusion}\label{summ}

In this article, the general features and characteristics of Bessel function series were investigated. Special emphasis was focused on Kapteyn series, which appear in many applications of theoretical physics and mathematics, such as radiation and optimization problems. In their original form, with the index of summation appearing in both the index and the argument of the Bessel function(s) involved, the convergence of such series is, in general, unclear. Therefore, it is appropriate and necessary to undertake every effort of rewriting such sums in terms of, at worst, infinite power series or (double) integrals, the convergence of which can be estimated more reliably. More importantly, in many cases it has proven possible to find closed analytical expressions for Kapteyn series of both the first and the second kind. This is indispensable for cases where Kapteyn series constitute only part of large mathematical expressions that have to be dealt with numerically.

However, the quest for closed-form expressions of Kapteyn series has mostly been limited to special cases that have arisen in specific problems such as those listed in Sec.~\ref{intro}. Often one notes that a similar procedure proves useful for different forms of Kapteyn series, even in such cases where the summation coefficients are considerably diverse. But no general answer has been found to date to the following problem: For which parameter regimes of the coefficients do the Kapteyn series have a closed-form expression?

It was the aim of the present article to shed some light on that question. Starting from the most general form of Kapteyn series of the second kind, i.\,e., involving the product of two Bessel functions, the Kapteyn series was decomposed into series over trigonometric functions or, more generally, algebraic expressions involving gamma functions. The ability to sum such series depends on the precise choice of the parameters. However, it has been shown that the likelihood for any analytical tractability is increased if both Bessel functions are of equal order, i.\,e., $(\alpha n+\beta)=(\gamma n+\epsilon)$ in equation~\eqref{eq:K2}.

Two specific examples with applications, for instance, in cosmic ray diffusion theories \citep{tau10:die,sha04:mhd,tau06:sta} were illustrated, where the summation coefficients are simply powers of the summation index. It has been shown that, depending very sensitively on the parameters chosen for the summation coefficients, power series with a finite/infinite number of terms are obtained, where the relative magnitude of each term can now easily be estimated.

Future work should, presumably, concentrate on the application of one or more of the above methods (or indeed combinations of the methods) to determine to the maximum extent possible the broadest range of conditions for summability of the general Kapteyn series. Of course, new methods for attempting such summations are welcome and it would be highly interesting to see any such ideas that would add to the capability to effect Kapteyn series summations. Another interesting question is to decide if for every Kapteyn series there exists an annihilator differential operator that can be used to obtain structural relations between series of different orders.

\ack
{\it The authors are thankful to the digitalization services of several online archives, without which the access to the work of mathematicians around the beginning of the 20\textsuperscript{th} Century would be considerably more difficult. The work of D. Dominici was partially supported by a Humboldt Research Fellowship for Experienced Researchers from the Alexander von Humboldt Foundation.}


\end{document}